\documentclass[conference]{IEEEtran}
\usepackage{booktabs}
\usepackage{multirow}
\IEEEoverridecommandlockouts
\usepackage{amsmath,amssymb,amsfonts}
\usepackage{algorithmic}
\usepackage{graphicx}
\usepackage{textcomp}
\usepackage{xcolor}
\usepackage[backend=bibtex,bibstyle=ieee,citestyle=numeric-comp]{biblatex}
\usepackage{caption} 
\usepackage{hyperref}
\def \partha [#1]{\textcolor{red}{Partha: #1}} 

\begin{document}

\title{ Representation Learning for Semantic Alignment of Language, Audio, and Visual  Modalities\\
%{\footnotesize \textsuperscript{*}note: Sub-titles are not captured in Xplore and
}
% Contrastive Language alignment Modality 

%\author{\IEEEauthorblockN{Samuel Lipping}
%\IEEEauthorblockA{\textit{Audio Research Group} \\
%\textit{Tampere University}\\
%Tampere, Finland\\
%samuel.lipping@tuni.fi}
%\and

%\IEEEauthorblockN{Parthasaarathy Sudarsanam}
%\IEEEauthorblockA{\textit{Audio Research Group} \\
%\textit{Tampere University}\\
%Tampere, Finland\\
%parthasaarathy.ariyakulamsudarsanam@tuni.fi}
%\and

%\IEEEauthorblockN{Konstantinos Drossos}
%\IEEEauthorblockA{\textit{Audio Research Group} \\
%\textit{Tampere University}\\
%Tampere, Finland\\
%konstantinos.drossos@tuni.fi}
%\and
%\IEEEauthorblockN{Tuomas Virtanen}
%\IEEEauthorblockA{\textit{Audio Research Group} \\
%\textit{Tampere University}\\
%Tampere, Finland\\
%tuomas.virtanen@tuni.fi}
%}

\author{
    \IEEEauthorblockN{ Parthasaarathy Sudarsanam, Irene Martín-Morató, Tuomas Virtanen}
    \IEEEauthorblockA{\textit{Audio Research Group, Tampere University, Tampere, Finland}}
   % \IEEEauthorblockA{\textit{}}
    \IEEEauthorblockA{\{parthasaarathy.ariyakulamsudarsanam, irene.martinmorato, tuomas.virtanen\}@tuni.fi}
}

\maketitle
\begin{abstract}

This paper proposes a single-stage training approach that semantically aligns three modalities - audio, visual, and text using a contrastive learning framework. Contrastive training has gained prominence for multimodal alignment, utilizing large-scale unlabeled data to learn shared representations. Existing deep learning approach for trimodal alignment involves two-stages, that separately align visual-text and audio-text modalities. This approach suffers from mismatched data distributions, resulting in suboptimal alignment. Leveraging the AVCaps dataset, which provides audio, visual and audio-visual captions for video clips, our method jointly optimizes the representation of all the modalities using contrastive training. Our results demonstrate that the single-stage approach outperforms the two-stage method, achieving a two-fold improvement in audio based visual retrieval, highlighting the advantages of unified multimodal representation learning.

\end{abstract}


\section{Introduction}

The increasing prevalence of multimodal content and large language models demands the development of models capable of jointly processing and understanding audio, visual (video frames without sound), and textual modalities. Contrastive learning techniques~\cite{CLIP_Radford2021, CLAP_Elizalde2023} have proven effective in bridging modality gaps. They achieve better semantic alignment across modalities compared to earlier deep learning approaches. For example, multimodal deep Boltzmann machines~\cite{NIPS2012_af21d0c9} relied on generative modeling,~\cite{joulin2016learning} demonstrated that CNNs trained to predict words from image captions could learn useful multimodal representations. Contrastive learning frameworks explicitly optimize for multimodal similarity achieving superior alignment. %While existing methods typically focus on semantically aligning pairs of these modalities, they fall short in capturing the interactions among all three. This limits development of systems capable of capturing the nuanced relationships inherent in multimodal data. Tackling this challenge is crucial for achieving more robust and contextually aware multimodal AI systems.

Multimodal representation learning has seen significant advancements in aligning two modalities. Models like ALIGN~\cite{jia2021scaling}, CLIP~\cite{CLIP_Radford2021}, Florence~\cite{yuan2021florence} achieve impressive results by aligning image-text pairs; and CLAP~\cite{CLAP_Elizalde2023} performs effectively in aligning audio-text pairs, using the contrastive learning frameworks. However, extending this success to trimodal alignment remains a challenge. Two-stage approaches, such as wav2CLIP~\cite{wu2022wav2clip}, first align visual and textual modalities in a shared embedding space and then introduce audio alignment in a second stage. While effective for certain tasks, this sequential process often leads to suboptimal audio-visual alignment, as the model is biased towards the initial visual-text relationship. Moreover, differing dataset distributions between the two stages amplify the alignment issues, making it difficult to fully capture the intricate interactions among audio, visual, and textual modalities.

\begin{figure}[]
\centering
\includegraphics[width=0.8\columnwidth]{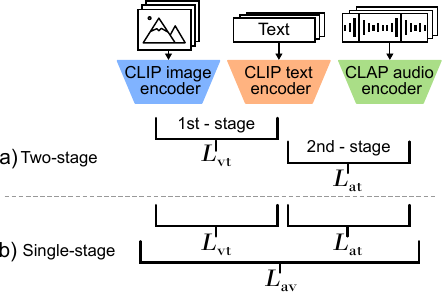}
\caption{Schematic representation of the two-stage reference method~(a) and the proposed single-stage method~(b). %$L_{\text{vt}}$, $L_{\text{at}}$, $L_{\text{av}}$ represent the contrastive losses between visual and text, audio and text and audio and visual modalities, respectively. 
Our proposed method aligns the representations from the three modalities by jointly optimizing the visual-text ($L_{\text{vt}}$), audio-text ($L_{\text{at}}$), and audio-visual ($L_{\text{av}}$) contrastive losses. % when training the encoders that learn the representations. 
}
\label{two-single-models}
\end{figure}

Trimodal representation learning is further hampered by the lack of datasets with modality-specific textual annotations for audio and visual data for the same video clip. Existing datasets, focus on visual-text~\cite{lin2014microsoft, krishna2017visual, thomee2016yfcc100m} or audio-text~\cite{fonseca2022fsd50k, drossos2020clotho, kim2019audiocaps, martin2021groundtruth} pairs, leaving a gap in capturing the complex relationships required for unified trimodal models. AudioCLIP~\cite{audioCLIP_Guzhov2022} attempts to bridge this gap by extending CLIP with an audio branch and aligning all three modalities using sound event classes from AudioSet~\cite{Gemmeke2017} as textual descriptions. However, these annotations are primarily sound-centric and lack visual context. These limitations highlight the need for unified frameworks and datasets that enable simultaneous learning across all three modalities. Recently, we released the AVCaps dataset~\cite{sudarsanam_2024_14536325} that contains modality-specific captions for audio, visual, and audiovisual modalities associated with video clips.

%Recently, we released the AVCaps %\footnote{The AVCaps dataset is available online at \href{https://huggingface.co/datasets/TUT-ARG/AVCaps}{Hugging Face} and \href{https://zenodo.org/records/14536325}{Zenodo}.} 
%dataset~\cite{sudarsanam_2024_14536325} that has modality-specific captions for audio, visual, and audiovisual modalities associated with video clips. Leveraging this dataset, in this work, we present SLAVA, a Single-stage Language Audio Visual Alignment model contrastively trained to semantically align audio, visual, and textual modalities effectively. We compare our approach against two-stage tri-modal alignment models and single-stage alignment models akin to Wav2CLIP and AudioCLIP respectively, demonstrating that our single-stage framework achieves superior tri-modal alignment.

This paper presents SLAVA, a Single-stage Language Audio Visual Alignment model contrastively trained to semantically align the three modalities effectively. Unlike previous methods that use either a two-stage approach~\cite{wu2022wav2clip} or a single-stage model with only audio captions~\cite{audioCLIP_Guzhov2022}, we leverage the AVCaps dataset to simultaneously utilize both audio and visual captions. This allows us to align all three modalities within a single-stage framework. Our results demonstrate that SLAVA achieves superior trimodal alignment, outperforming existing methods. % Recently, we released the AVCaps dataset~\cite{sudarsanam_2024_14536325} that contains modality-specific captions for audio, visual, and audiovisual modalities associated with video clips. 
%Leveraging this dataset, we compare our approach against two-stage trimodal alignment models and single-stage trimodal models akin to Wav2CLIP and AudioCLIP respectively, demonstrating that our single-stage framework achieves superior trimodal alignment.

%The remainder of this paper is organized as follows. Section~\ref{related_works} reviews related work in multimodal alignment. Section~\ref{experiments} describes the dataset, reference and proposed methods for tri-modal alignment. Section~\ref{evaluation} outlines the evaluation metrics, and Section~\ref{results} presents and analyzes the experimental results. Finally, Section~\ref{conclusion} summarizes our findings.

\section{Related works}\label{related_works}

\subsection{Contrastive Language-Image Pretraining (CLIP)}

The CLIP model~\cite{CLIP_Radford2021} is a powerful contrastive learning framework designed to align textual and visual modalities. Trained on large-scale of 400M image-text pairs, it enables the model to learn meaningful correspondences between natural language descriptions and images. However, its reliance on object-centric captions in the training data limits its capacity to capture event-level or scene-level relationships. Additionally, since the text encoder was trained solely with captions derived for static images, it has not been exposed to audio-related language. This restricts CLIP's ability to generalize to tasks involving audio-visual contexts or to represent relationships beyond the visual modality.

\subsection{Contrastive Language-Audio Pretraining (CLAP)}

CLAP~\cite{CLAP_Elizalde2023} uses contrastive learning to align textual and audio modalities. Training on audio-text pairs, CLAP enables associations between natural language descriptions and diverse audio inputs, such as environmental sounds or speech. CLAP is trained with a relatively smaller dataset of 128K audio clips and their associated textual description sourced from various audio captioning datasets. Similar to the limitations of CLIP, The CLAP text encoder has not been exposed to vision specific language during training, constraining its ability to represent visual relationships. 

\subsection{Wav2CLIP}
%The limitations of CLIP and CLAP highlight the need for models that can learn trimodal representations. 
Wav2CLIP~\cite{wu2022wav2clip} uses a two-stage approach for audio-visual-textual alignment. It involves training a visual-text semantic alignment model in the first stage. In the second stage, one of the branches is either frozen or fine-tuned along with an audio-text or audio-visual dataset to achieve trimodal alignment. A common issue with this method arises from using datasets from different sources for the two stages, resulting in data distribution mismatches that lead to suboptimal alignment.

\subsection{AudioCLIP}

In AudioCLIP~\cite{audioCLIP_Guzhov2022}, the authors proposed a single stage model by utilizing AudioSet~\cite{Gemmeke2017}, a dataset that includes video clips associated with sound events, as a foundation for its trimodal training. AudioCLIP extends CLIP by integrating an audio encoder to the CLIP framework to create a unified model. While this trimodal approach represents significant progress, it has certain limitations. The textual descriptions used for aligning the three modalities are derived from Audioset sound-event classes, resulting in annotations that are primarily focused on sound-event-specific information. This focus limits the model's ability to generalize to tasks requiring balanced, scene-level, or context-rich representations across audio-visual modalities.

%The limitations of these models reinforces the necessity for datasets and models capable of learning tri-modal audio-visual-textual representations. 

\section{METHODS}\label{experiments}
\begin{table*}[ht]
    \centering
    \renewcommand{\arraystretch}{1.2} % Increase row height for better readability
    \begin{tabular}{l l l l l}
        \toprule
        \textbf{Methods} & \textbf{Stages} & \textbf{Trainable Layers} & \textbf{Training data} & \textbf{Objective Function} \\
        \midrule
        \multirow{2}{*}{Wav2CLIP-style}  
            & Stage 1 & Visual and text projection & Visuals and visual captions & $L_{\text{vt}}$ \\
            & Stage 2 & Audio and text projection & Audio and audio captions & $L_{\text{at}}$ \\
        \midrule
       AudioCLIP-style  & Single stage & Audio, visual, and text projection & Audios, visuals, audio captions & $L_{\text{av}} + L_{\text{vt}} + L_{\text{at}}$ \\
        \midrule
        SLAVA\textsubscript{A\&V} (ours) & Single stage & Audio, visual, and text projection & Audios, visuals, audio captions, visual captions & $L_{\text{av}} + ( L_{\text{vt}} \textnormal{ or } L_{\text{at}} )$ \\
        \midrule
        SLAVA\textsubscript{AV} (ours) & Single stage & Audio, visual, and text projection & Audios, visuals, audio-visual captions & $L_{\text{av}} + L_{\text{vt}} + L_{\text{at}}$ \\
        \bottomrule
    \end{tabular}
    \caption{Overview of the reference and proposed models, training stages, trainable layers, datasets, and loss functions. $L_{\text{av}}$, $L_{\text{at}}$, and $L_{\text{vt}}$ represent the audio-visual, audio-text, and visual-text contrastive losses respectively.}
    \label{tab:model_comparison}
\end{table*}

\subsection{Dataset}

We use our recently released AVCaps~\cite{sudarsanam_2024_14536325} dataset. The dataset comprises 2,061 videos with a total duration of 28.8 hours. Each video is annotated with up to five crowdsourced captions for audio, visual, and audio-visual modalities. In addition, each video includes three audio-visual captions generated using GPT-4, leveraging the modality-specific captions. This uniqueness of AVCaps dataset %, containing modality-specific captions for the same video clip,
enables a comprehensive study of training strategies for aligning audio, visual, and textual modalities. % Further details about the dataset can be found in~\cite{sudarsanam_2024_14536325}.

\subsection{Overview of the methods}

In all our experiments, we used pretrained CLIP image and text encoders and CLAP audio encoder and finetuned their projection layers to obtain latent representations of visual, textual, and audio inputs. %Specifically, we used the CLIP image and text encoders and the CLAP audio encoder. % The feature extraction layers of all encoders were frozen, while the projection layers were finetuned on the AVCaps dataset.
The CLIP image encoder processes an input video of $M$ frames individually to produce a representation of size $\mathbb{R}^{M\times768}$, while the text encoder generates a $\mathbb{R}^{512}$ representation for a given textual input. The CLAP audio encoder takes $N$ chunks of 10-second audio inputs and returns a representation of size $\mathbb{R}^{N\times768}$.

The encoded representations are passed through their respective projection layers. The CLIP image and CLAP audio projection layers, each a linear layer with 512 neurons, convert $\mathbb{R}^{M\times768}$ and $\mathbb{R}^{N\times768}$ to $\mathbb{R}^{M\times512}$ and $\mathbb{R}^{N\times512}$, respectively.
%The CLIP image projection converts the $\mathbb{R}^{M\times768}$ representation to $\mathbb{R}^{M\times512}$. The CLAP audio projection layer similarly transforms the $\mathbb{R}^{N\times768}$ representation to $\mathbb{R}^{N\times512}$. 
The CLIP text projection layer is also a linear layer with 512 neurons and hence keeps the size of the text representation unchanged. 
To make the representations of the three modalities the same size, we compute the average of the audio and visual features along the time axis, resulting in a final representation of $\mathbb{R}^{512}$. The projection layers, which learn the joint multimodal representation are finetuned using the InfoNCE~\cite{oord2018representation} contrastive loss.

%To ensure a fair comparison with our proposed approach, we designed two reference methods using the AVCaps dataset. The first follows a two-stage training paradigm, similar to Wav2CLIP, where visual-text and audio-text alignment are learned in separate stages. The second adopts a single-stage training approach, inspired by AudioCLIP, where all three modalities are trained jointly. 
Table~\ref{tab:model_comparison} provides an overview of the reference methods alongside our proposed approaches, highlighting the trainable layers, the AVCaps data partitions used, and the loss functions employed. These methods are further detailed in the following subsections.

\subsection{Reference Methods}
\subsubsection{\textbf{Wav2CLIP-style two-stage model}}
In this approach, we first align the visual and textual representations, followed by the alignment of audio and textual representations in the next stage. We use the AVCaps dataset, which provides audio and visual captions for the same videos, to address data distribution mismatches in conventional two-stage models. We begin by fine-tuning the projection layers of a pre-trained CLIP model, using the visual content from the videos and their crowdsourced visual captions. For the second stage, we finetune the audio projection layers of a pre-trained CLAP model alongside the textual projection layers of the CLIP text encoder from stage one. Alternatively, we experimented with keeping the text projection layers frozen, only fine-tuning the audio projection layers in the second stage. %By ensuring that the datasets used in both stages come from the same distribution, this approach mitigates distribution mismatch. 
This approach is similar to the training paradigm used in Wav2CLIP. In this approach, we align the audio and visual embeddings indirectly through shared textual representations, which may result in suboptimal performance for audio-visual tasks. \newline
%Figure ?? illustrates our two-stage training process. YESS! Figure is needed

\subsubsection{\textbf{AudioCLIP-style single-stage model}}

As an additional reference, we developed a single-stage model inspired by the AudioCLIP framework. In this model, we fine-tuned the pre-trained CLIP projection layers and CLAP audio projection layers using the visuals, audios, and the audio captions from the AVCaps dataset using pairwise contrastive losses for all three modalities. This approach mirrors the AudioCLIP training process, which utilized the AudioSet, as their captions predominantly focus on sound events. By aligning the three modalities directly in a single-stage framework, this method provides an additional benchmark for evaluating the performance of our proposed trimodal alignment approach. The single-stage approach addresses the indirect audio-visual alignment limitation of the two-stage approach. However, its effectiveness is limited by the fine-tuning dataset, where the textual descriptions cover only a single modality, in this case audio.

\subsection{Proposed methods}
\subsubsection{\textbf{SLAVA with audio and visual captions}}

%The two-stage approach indirectly aligns audio and visual embeddings through the shared textual representations.
%The two-stage approach lacks a direct mechanism to align audio and visual embeddings, as these modalities are connected only through the shared textual representations. 
%This indirect alignment can lead to suboptimal performance for audio-visual tasks. The single-stage approach addresses this limitation by directly aligning all three modalities using pairwise contrastive losses. However, its effectiveness is limited by the fine-tuning dataset, where the textual descriptions cover only a single modality, in this case audio.

Our proposed SLAVA model uses both audio and visual captions for single-stage training to enhance multimodal alignment. In this setup, the model benefits from both auditory and visual information present in the textual captions, enhancing the alignment between all three modalities. %Our approach addresses the limitations of our reference methods by: 1) enabling explicit alignment of both audio and visual modalities, and 2) ensuring that the textual descriptions comprehensively cover both audio and visual modalities. 
Figure~\ref{two-single-models}b shows our proposed SLAVA model. For each minibatch, we choose corresponding audios, visuals, and randomly either audio captions or visual captions from the AVCaps dataset. When audio captions are used, the loss includes audio-text and audio-visual contrastive terms. Conversely, when visual captions are used, the loss comprises visual-text and audio-visual contrastive terms. We refer to this model as SLAVA\textsubscript{A\&V} denoting that it was trained with audio and visual captions. The total loss $L_{\text{total}}$ in this case is given by

\begin{equation}
    L_{\text{total}} = L_{\text{av}} + (L_{\text{at}} \text{ or } L_{\text{vt}})
    \label{eq:loss_function_1}
\end{equation}

\noindent where, $L_{\text{av}}$, $L_{\text{at}}$, and $L_{\text{vt}}$ represent the audio-visual, audio-text, and visual-text contrastive losses respectively. Our approach addresses the limitations of our reference methods by 1) enabling explicit alignment of both audio and visual modalities, and 2) ensuring that the textual descriptions comprehensively cover both audio and visual modalities.%In this way AVCaps enables us to address the limitations of both our references methods. \newline

%For every minibatch either $L_{\text{at}}$ or $L_{\text{vt}}$ is used depending on whether the textual inputs are audio captions or visual captions. 

% Additionally, we experimented with an alternative strategy to enhance cross-modal alignment by introducing an extra loss term. Specifically, when the input caption is audio, we add a visual-text contrastive loss, and vice versa. This setup enables a direct comparison between single-stage and two-stage strategies, using the same input texts in both cases.
\subsubsection{\textbf{SLAVA with audio-visual captions}}
In another set of experiments, we used the audios, visuals and the LLM-generated audio-visual captions. We refer to this model as SLAVA\textsubscript{AV} denoting that it was trained on audio-visual captions. For these experiments, we designed two training configurations. In the first configuration, training was performed using two contrastive losses $L_{\text{at}}$ and $L_{\text{vt}}$. This approach allows us to compare how well the captions containing both audio and visual information could indirectly improve audio-visual alignment without explicitly applying $L_{\text{av}}$. In this configuration, the loss function is given by

\begin{equation}
    L_{\text{total}} = L_{\text{at}} + L_{\text{vt}}
    \label{eq:loss_function_2}
\end{equation}

In the second configuration, we included the $L_{\text{av}}$, thereby explicitly aligning all three modalities. The loss function is given by

\begin{equation}
    L_{\text{total}} = L_{\text{av}} + L_{\text{at}} + L_{\text{vt}}
    \label{eq:loss_function_3}
\end{equation}

%We refer to this model as SLAVA\textsubscript{AV} denoting that it was trained on audio-visual captions. These configurations allow for a thorough examination of the impact of audio-visual captions on both indirect and direct trimodal alignment.

\section{EVALUATION} \label{evaluation}
\begin{table*}[]
    \centering
    \renewcommand{\arraystretch}{1.5} % Increase row height for better spacing
    \begin{tabular}{|c|c|c|c|c|c|c|c|}
    \hline
        \textbf{Retrieve} & \textbf{Based On} & 
        \multicolumn{2}{c|}{\textbf{Wav2CLIP-style (2 stage)}} & 
        \textbf{AudioCLIP-style} & 
        \textbf{SLAVA\textsubscript{A\&V}} & 
        \multicolumn{2}{c|}{\textbf{SLAVA\textsubscript{AV} (ours)}} \\
        \cline{3-4} \cline{7-8}
         &  & \textbf{Frozen Text} & \textbf{Trainable Text} & \textbf{(single stage)} & \textbf{(ours)} & $L\textsubscript{at} + L\textsubscript{vt}$ & $L\textsubscript{av} +L\textsubscript{at} + L\textsubscript{vt}$ \\
    \hline
        Visual & Visual Captions & 
        0.84 & 0.82 & 0.69 & 0.81 & 0.83 & 0.81 \\
        Visual & Audio Captions & 
        0.26 & 0.29 & 0.29 & 0.30 & 0.29 & 0.29 \\
        Visual & Audio-Visual Captions & 
        0.86 & 0.86 & 0.75 & 0.85 & 0.89 & 0.86 \\
    \hline
        Audio & Audio Captions & 
        0.35 & 0.40 & 0.36 & 0.37 & 0.33 & 0.34 \\
        Audio & Visual Captions & 
        0.27 & 0.29 & 0.26 & 0.32 & 0.32 & 0.33 \\
        Audio & Audio-Visual Captions & 
        0.41 & 0.41 & 0.37 & 0.40 & 0.44 & 0.45 \\
    \hline
        Visual & Audio & 
        0.22 & 0.27 & 0.45 & 0.52 & 0.40 & 0.50 \\
    \hline
    \end{tabular}
    \caption{\textbf{Recall@10} retrieval performance across various models and configurations on the AVCaps test split.} %\textbf{Reference method - I} refers to a two-stage approach with two configurations: frozen and trainable text projections in the second stage. \textbf{Reference method - II} represents a single-stage model trained using audio captions.} %\textbf{Proposed method - I} incorporates audio and visual captions in a single-stage setup. \textbf{Proposed method - II} utilizes audio-visual captions with two configurations: one without the audio-visual contrastive loss (2 losses) and another with the audio-visual contrastive loss (3 losses).}
    \label{tab:comparison_table}
\end{table*}
To evaluate our reference and proposed methods, we performed a series of retrieval tasks designed to assess bimodal and trimodal alignment. %As detailed in Section~\ref{experiments}, all our models utilize pre-trained CLIP and CLAP encoders, whose projection layers are finetuned on the training split of the AVCaps dataset.
All our models were finetuned on the AVCaps training split for $20$ epochs, using the AdamW optimizer with a learning rate of $1e-5$ and a weight decay of $0.1$ using the InfoNCE loss~\cite{oord2018representation}. The best model is selected based on its performance on the validation split of AVCaps.

We evaluated all our models on the test split of AVCaps on several retrieval tasks with the widely used recall@10 metric. In a retrieval task, a textual caption or an audio signal serves as the input query, which is compared against all items in the database to retrieve the top-$10$ most similar results. The recall@10 metric calculates the proportion of relevant items among the top-$10$ retrieved results by the total of items in the dataset.

%One modality caption is used as query, then the similarity between the query and items (visual or audios) is calculated, the top $10$ most similar are retrieved.

%It checks whether the target is present in the top $10$ predictions of the model. 

The bimodal retrieval tasks evaluate the model's ability to retrieve one modality using queries from another, such as retrieving audio with audio captions or visuals with visual captions. For crossmodal retrieval, we investigate tasks like audio captions based visual retrieval and visual captions based audio retrieval, which require the model to align representations across different modalities. We also evaluate audio-visual captions based audio and visual retrieval. Additionally, to compare our models in terms of audio-visual similarities, we analyze the performance of audio based visual retrieval to assess how the models align and transfer knowledge between these two modalities. These tasks provide a comprehensive evaluation of the model's ability in creating a shared multimodal representation for all three modalities involved.

\section{RESULTS} \label{results}

%In this section, we present the results of all our retrieval experiments.

Table~\ref{tab:comparison_table} shows the retrieval performance of all our models on the AVCaps test split evaluated using the recall@10 metric. Wav2CLIP-style results show the retrieval performance of our reference two-stage model in two configurations: frozen and trainable text projection layers in stage 2. In the first approach, the text encoder is trained only with visual captions and CLIP training data, which mainly contain vision-specific language. When audio captions are introduced in stage 2, the frozen text encoder produces suboptimal representations, constraining the audio representation to align with these suboptimal text embeddings. This results in a weaker audio-visual alignment. In contrast, updating the text projection layer in stage 2 enables the audio encoder to learn better representations of the input audio, resulting in better alignment. %Additionally, fine-tuning the text projection layers based on audio representations in stage 2 causes a slight decrease in the performance of visual retrieval using visual captions compared to the first approach. However, the performance of audio retrieval based on audio captions, as well as the cross-modal retrieval tasks, all show improvements in the second approach.
Finetuning text projection layers slightly lowers visual retrieval with visual captions but improves audio retrieval and other crossmodal tasks.
%While finetuning text projection layers slightly reduces visual retrieval performance with visual captions, it enhances audio retrieval and other crossmodal tasks.

AudioCLIP-style results present the performance of our reference single-stage model, which aligns all three modalities using audio captions. It improves the  audio-visual alignment, with audio-based visual retrieval increasing to 0.45 compared to 0.27 of the two-stage approach. However, the visual retrieval scores drop significantly in comparison to the two-stage approach. This is because the captions contain audio-specific information which shifts the distribution of the text encoder away from its original visual-centric pretraining, reducing its ability to capture visual features.

SLAVA\textsubscript{A\&V} results summarize the performance of our proposed single-stage model trained on audio and visual captions. It uses  audio-visual contrastive loss combined with either audio-text or visual-text contrastive loss, based on the textual input in the mini-batch. By directly incorporating audio-visual contrastive loss, the single-stage approach improves audio-based visual retrieval, increasing recall@10 to 0.52 from 0.27, while maintaining performance across other retrieval tasks. These results clearly establish that single-stage trimodal alignment using both audio and visual captions enabled by AVCaps dataset create a robust multimodal representation compared to the traditional two-stage approach. 

Finally, we present the results of our SLAVA\textsubscript{AV} model, a single-stage model trained with audio-visual captions. The first approach, using only audio-text and visual-text contrastive losses, is designed for comparison with the two-stage model lacking explicit audio-visual alignment. Despite this, our proposed model achieves a recall@10 of 0.40 for audio-based visual retrieval, compared to 0.27 for the two-stage model. In the second approach, the model incorporates an additional audio-visual contrastive loss alongside audio-text and visual-text losses, enabling explicit alignment across the modalities. This model outperforms the reference single stage model on almost all the retrieval tasks across modalities.

%\begin{table}[ht]
%    \centering
%    \renewcommand{\arraystretch}{1.5} % Increase row height for better spacing
%    \begin{tabular}{|c|c|c|}
%    \hline
%        \textbf{Retrieve } & \textbf{Based On} & \textbf{R@10} \\
%        \hline
%%        Visual & Audio & 0.50\\
 %   \hline
 %   \end{tabular}
 %   \caption{Retrieval performance of the audio-visual alignment model}
 %   \label{tab:audio-visual_alignment}
%\end{table}

We also trained an audio-visual alignment model using the CLIP image encoder and the CLAP audio encoder, with audio and visuals from the AVCaps training split.  This model achieved a score of 0.50 in audio-based visual retrieval on the test split. Our proposed SLAVA models match this performance while also effectively aligning the textual modality with both the audio and visual modalities.

%with the results presented in Table~\ref{tab:audio-visual_alignment}. Notably, this model achieved a score of 0.50, demonstrating that our tri-modal alignment approach can match this performance while also effectively aligning the textual modality with the other two modalities.

\section{Conclusion} \label{conclusion}

In this paper, we proposed SLAVA, a single-stage approach for multimodal alignment that jointly optimizes audio, visual, and textual representations using contrastive training. Our method addresses the limitations of two-stage approaches, where separate training of visual-text and audio-text models often leads to suboptimal alignment due to mismatched data distributions. Using modality-specific captions in the AVCaps dataset, we demonstrated that our approach significantly improves alignment across modalities. Evaluation results on a variety of retrieval tasks, show that our single-stage method outperforms the traditional two-stage approach, achieving a recall@10 score of 0.52 in audio-based visual retrieval compared to 0.27 of the reference model. These findings highlight the advantages of AVCaps dataset and a unified framework for multimodal representation learning, demonstrating its potential for advancing tasks such as audio-visual captioning, video understanding, and multimodal question answering.

\section*{Acknowledgment}

The funding for this work is supported by Jane and Aatos Erkko Foundation through the CONVERGENCE of Humans and Machines project.

\printbibliography

\end{document}